\documentclass[aps,prx,twocolumn,showpacs,superscriptaddress,amsmath,amssymb]{revtex4-1}

\usepackage{graphicx}
\usepackage[colorlinks=true,linkcolor=blue,citecolor=blue,urlcolor=blue]{hyperref}
\usepackage{bbold}
\usepackage{gensymb}

\AtBeginDocument{%
    \newwrite\bibnotes
    \def\bibnotesext{Notes.bib}
    \immediate\openout\bibnotes=\jobname\bibnotesext
    \immediate\write\bibnotes{@CONTROL{REVTEX41Control}}
    \immediate\write\bibnotes{@CONTROL{%
    apsrev41Control,author="08",editor="1",pages="0",title="0",year="1"}}
     \if@filesw
     \immediate\write\@auxout{\string\citation{apsrev41Control}}%
    \fi
}%

\def \D {\Delta}
\def \e {\epsilon}
\def \ve {\varepsilon}

\def \L {\Lambda}

\def \O {\Omega}

\def \s {\sigma}

\def \p {\partial}

\def \gradk {\nabla_{\mbf k}}

\def \curlk {\nabla_{\mbf k}\times}

\newcommand{\intv}[1]{\int_{\mbf #1}}

\def \la {\langle}
\def \ra {\rangle}
\def \fr {\frac}

\newcommand{\bra}[1]{\la#1|}
\newcommand{\ket}[1]{|#1\ra}
\newcommand{\braket}[3]{\la#1|#2|#3\ra}
\newcommand{\innp}[2]{\la#1|#2\ra}

\newcommand{\epvl}[1]{\la#1\ra}

\def \Tr {\mathrm{Tr}}
\def \bece {\begin{center}}
\def \ence {\end{center}}
\def \beeq {\begin{equation}}
\def \eneq {\end{equation}}
\def \beal {\begin{aligned}}
\def \enal {\end{aligned}}
\def \bega {\begin{gathered}}
\def \enga {\end{gathered}}
\def \benu {\begin{enumerate}}
\def \ennu {\end{enumerate}}
\def \beit {\begin{itemize}}
\def \enit {\end{itemize}}
\def \bede {\begin{description}}
\def \ende {\end{description}}
\def \betb {\begin{tabular}}
\def \entb {\end{tabular}}
\def \bear {\begin{array}}
\def \enar {\end{array}}

\def \mbf {\mathbf}
\def \mbb {\mathbb}

\def \mca {\mathcal}

\def \bsb{\boldsymbol}
\def \txt {\text}

\newcommand{\comment}[1]{}

\begin{document}


\title{Band geometry from position-momentum duality at topological band crossings}

\author{Yu-Ping Lin}
\affiliation{Department of Physics, University of Colorado, Boulder, Colorado 80309, USA}
\author{Wei-Han Hsiao}
\affiliation{Independent Researcher, Chicago, Illinois, USA}

\date{\today}

\begin{abstract}
We show that the position-momentum duality offers a transparent interpretation of the band geometry at the topological band crossings. Under this duality, the band geometry with Berry connection is dual to the free-electron motion under gauge field. This identifies the trace of quantum metric as the dual energy in momentum space. The band crossings with Berry defects thus induce the dual energy quantization in the trace of quantum metric. For the $\mathbb Z$ nodal-point and nodal-surface semimetals in three dimensions, the dual Landau level quantization occurs owing to the Berry charges. Meanwhile, the two-dimensional (2D) Dirac points exhibit the Berry vortices, leading to the quantized dual axial rotational energies. Such a quantization naturally generalizes to the three-dimensional (3D) nodal-loop semimetals, where the nodal loops host the Berry vortex lines. The $\mathbb Z_2$ monopoles bring about additional dual axial rotational energies, which originate from the links with additional nodal lines. Nontrivial band geometry generically induces finite spread in the Wannier functions. While the spread manifest quantized lower bounds from the Berry charges, logarithmic divergences occur from the Berry vortices. The band geometry at the band crossings may be probed experimentally by a periodic-drive measurement.
\end{abstract}

\maketitle

\section{Introduction}

\begin{figure}[t]
\centering
\includegraphics[scale = 1]{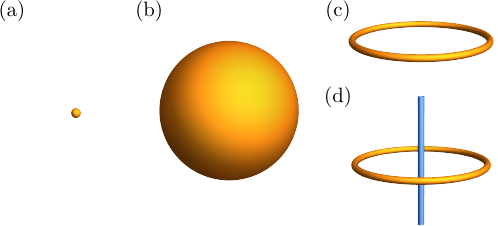}
\caption{\label{fig:nodal} Band crossings in the 3D Brillouin zone. (a) Nodal point. (b) Nodal surface. (c) $\mbb Z_2$-trivial nodal loop. (d) $\mbb Z_2$-nontrivial nodal loop.}
\end{figure}

The modern study of gapless topological systems has uncovered various unconventional band crossings with fascinating phenomena. In two dimensions, the band crossings may take the form of nodal points \cite{castroneto09rmp,lan11prb} and nodal lines \cite{li18prb}. Meanwhile, in three dimensions there exist nodal points \cite{armitage18rmp,bradlyn16sc,isobe16prb,tang17prl,boettcher20prl,rao19n,sanchez19n,schroter19np,takane19prl,lv19prb,Schroter20sc}, nodal lines \cite{armitage18rmp,carter12prb,weng15prx,fang15prb,liang16prb,zhao17prl,bzdusek17prb,ahn18prl}, and nodal surfaces \cite{liang16prb,bzdusek17prb,turker18prb,wu18prb} (Fig.~\ref{fig:nodal}). These band crossings can occur in the solid-state semimetals \cite{armitage18rmp}, topological superconductors \cite{sigrist91rmp,nandkishore16prb,agterberg17prl,brydon18prb,venderbos18prx,link20prl}, and spin liquids \cite{savary16rpp}. Recent studies of synthetic quantum matter present another set of platforms for the band crossings, including the ultracold atomic systems \cite{cooper19rmp,song19np}, photonic systems \cite{ozawa19rmp}, and superconducting circuits \cite{tan18prl,tan19prlweyl}. Distinct band-topology classifications are defined based on the symmetries and the band-crossing structures \cite{chiu16rmp,bzdusek17prb}. For the gapless topological systems, the topological classifications involve the $\mbb Z$ and $\mbb Z_2$ classifications. Various types of integer topological invariants are proposed as the indicators of these classifications. For example, a three-dimensional (3D) $\mbb Z$ nodal point [Fig.~\ref{fig:nodal}(a)] carries a Berry monopole, leading to a quantized Chern number under an enclosing-surface integration \cite{xiao10rmp}. The inflation of a $\mbb Z$ nodal point may realize a $\mbb Z$ nodal surface [Fig.~\ref{fig:nodal}(b)] \cite{agterberg17prl,brydon18prb}, where the Berry-charge scenario still applies \cite{turker18prb}. A 3D nodal line in a combined parity and time-reversal $PT$-symmetric system can carry a $\mbb Z_2$ monopole \cite{fang15prb,zhao17prl}. The according topological invariant corresponds to the linking number with additional nodal lines [Fig.~\ref{fig:nodal}(d)] \cite{ahn18prl}.

The band crossings also bring about nontrivial quantum geometry of the bands \cite{lin21prb}. The band geometry is characterized by the quantum metric, which measures the state variation under momentum change \cite{provost80cmp,page87pra,anandan90prl}. The components of quantum metric may be probed experimentally, for example, by a periodic drive \cite{ozawa18prb,chen20ax}. Recent studies uncovered various manifestations of the quantum metric. The integrated trace of quantum metric defines a lower bound of the spread of Wannier functions \cite{marzari97prb,matsuura10prb,marzari12rmp,lee14prb,chen17prb}, which can be measured directly in the experiments \cite{bleu18prb,ozawa18prb,asteria19np,klees20prl,yu19nsr,tan19prl,gianfrate20n}. The finite spread can trigger anomalous superfluid stiffness on flat bands, leading to the geometric enhancement of flat-band superconductivity  \cite{peotta15nc,liang17prb,hu19prl,xie20prl,lin20prr,lin21prb}. It can also lead to finite current noise even in the insulating phase \cite{neupert13prb}. Other works adopted the quantum metric as an indicator of phase transitions \cite{zanardi07prl,ma10prb,kolodrubetz13prb}, fractional Chern insulators \cite{roy14prb,claassen15prl,lee17prb}, excitons \cite{srivastava15prl}, and orbital susceptibilities \cite{gao15prb,piechon16prb}.

While the bulk of the literature focused on the extrinsic manifestations of the quantum metric, its intrinsic character has not received sufficient investigation. This direction was explored recently in the context of nodal-point semimetals. The trace of quantum metric has received a transparent interpretation in the chiral multifold semimetals \cite{lin21prb}. Under the position-momentum duality, the trace of quantum metric is dual to the kinetic energy on the Haldane sphere \cite{haldane83prl}, thereby acquiring a dual Landau level quantization from the Berry monopole. On the other hand, the integrated determinant of quantum metric over an enclosing surface was proposed as a measure of the Berry defect \cite{palumbo18prl,hwang21prb}. These results exemplified the nature of the quantum metric at the nodal points. A natural question then arises as whether these scenarios are applicable to the other topological systems.

In this work, we show that the position-momentum duality \cite{lin21prb} offers a unified framework of the band geometry at the topological band crossings. Under this duality, the band geometry with Berry connection is dual to the free-electron motion under gauge field. This identifies the trace of quantum metric as the dual energy in momentum space. The band crossings with Berry defects thus induce the dual energy quantization in the trace of quantum metric. For the 3D $\mbb Z$ nodal-point and nodal-surface semimetals, the dual Landau level quantization occurs owing to the Berry charges. Meanwhile, the two-dimensional (2D) Dirac points exhibit the Berry vortices, leading to the quantized dual axial rotational energies. Such a quantization naturally generalizes to the 3D nodal-loop semimetals, where the nodal loops host the Berry vortex lines. The $\mbb Z_2$ monopoles bring about additional dual rotational energies, which originate from the links with additional nodal lines. The spread of Wannier functions manifests quantized lower bounds from the Berry charges, while logarithmic divergences occur from the Berry vortices. Our duality-based analysis paves the way for advanced comprehension of nontrivial band geometry at the topological band crossings.

\section{Position-momentum duality}

We begin with an introduction to the band geometry. Consider a band with the Bloch eigenstate $\ket{u_{\mbf k}}$, which generically evolves under the variation of momentum $\mbf k$. Such an evolution manifests the variations in the phase and the state. A quantitative measure is provided by the quantum geometric tensor \cite{berry89}
\beeq
T_{ab\mbf k}=\braket{\p_{k_a}u_{\mbf k}}{(1-\ket{u_{\mbf k}}\bra{u_{\mbf k}})}{\p_{k_b}u_{\mbf k}},
\eneq
with the spatial indices $a,b=x,y,z$. The real part $g_{ab\mbf k}=\txt{Re}[T_{ab\mbf k}]$ is the quantum or Fubini-Study metric \cite{provost80cmp,page87pra,anandan90prl}, which captures the quantum distance under state variation $1-|\innp{u_{\mbf k}}{u_{\mbf k+d\mbf k}}|^2=g_{ab\mbf k}dk_adk_b$. On the other hand, the imaginary part measures the phase variation and the according Berry flux $B_{a\mbf k}=-\e_{abc}\txt{Im}[T_{bc\mbf k}]$ \cite{berry84rspa,xiao10rmp}. Define the Berry connection $\mbf A_{\mbf k}=\braket{u_{\mbf k}}{i\gradk}{u_{\mbf k}}$ as a momentum-space gauge field under nontrivial band geometry \cite{xiao10rmp}. The quantum metric and the Berry flux become \cite{shankar17chp}
\beeq
g_{ab\mbf k}=\fr{1}{2}\braket{u_{\mbf k}}{\{r_a,r_b\}}{u_{\mbf k}},\quad
\mbf B_{\mbf k}=\curlk\mbf A_{\mbf k},
\eneq
where the position $\mbf r=i\gradk-\mbf A_{\mbf k}$ is a momentum-space covariant derivative. The effect of Berry flux as a momentum-space ``magnetic field'' has been studied extensively \cite{xiao10rmp}. Here we focus on the quantum metric and search for useful indications to the band geometry.

The individual components of the quantum metric are usually complicated and hard to interpret. Nevertheless, the trace of quantum metric takes a profound form \cite{lin21prb}
\beeq
\label{eq:trg}
\Tr g_{\mbf k}=\epvl{|\mbf r|^2},
\eneq
where the expectation value $\epvl{\cdots}=\braket{u_{\mbf k}}{\cdots}{u_{\mbf k}}$ of momentum-space Laplacian $|\mbf r|^2$ is measured. Significantly, this quantity captures the ``dual energy'' of a free particle in momentum space, which carries a half-unity ``dual mass'' and experiences the Berry connection. We thus establish a position-momentum duality between the band geometry and the free-particle motion under gauge field. The duality presents a feasible solution to the interpretation of the band geometry. If the duality leads to a simple free-particle model with well-studied solution, the trace of quantum metric can be understood under a direct inference.

A similar concept of position-momentum duality was proposed previously in the contexts of harmonic traps \cite{wu11cpl,price14prl,ozawa16prb} and fractional Chern insulators \cite{claassen15prl,lee17prb}. These works adopted the confinement potential $V(\mbf r)\sim |\mbf r|^2$ to insert the band-geometry contribution into the actual electronic energy. In comparison, our analysis directly identifies the trace of quantum metric as the true dual energy in momentum space. This deepens the comprehension of the band geometry as the dual free-particle theory of general bands.

\section{Dual energy quantization in band geometry}

We now introduce how the position-momentum duality operates on the nontrivial band geometry. Here we focus on the gapless topological systems, where approximate rotation symmetries can occur near the band crossings. Under such symmetries, simple models become available to the dual free-particle theories. A band crossing can host a defect of Berry connection and induce nontrivial band geometry. We will show that this feature is encoded in the ``dual energy quantization'' in the trace of quantum metric. This scenario is applicable to the $\mbb Z$ and $\mbb Z_2$ gapless topological systems with the Berry charges and Berry vortices.

\subsection{Berry charges: Dual Haldane spheres}

\begin{figure}[t]
\centering
\includegraphics[scale = 1]{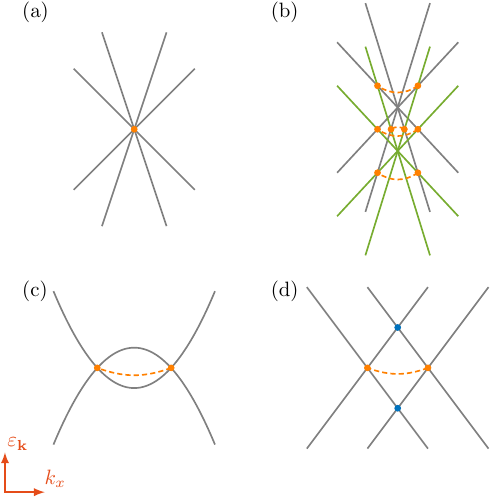}
\caption{\label{fig:band} The band structures in the minimal models for the (a) nodal-point, (b) nodal-surface, (c) $\mbb Z_2$-trivial nodal-loop, and (d) $\mbb Z_2$-nontrivial nodal-loop semimetals. The band crossings are indicated by the same colors as in Fig.~\ref{fig:nodal}. The points belonging to the same band crossing are connected by a dashed curve.}
\end{figure}

We first study the 3D $\mbb Z$ gapless topological systems with the Berry charges. As a paradigmatic example, we present the analysis of the chiral multifold semimetals (CMS) \cite{lin21prb}, which constitute a large family of $\mbb Z$ nodal-point semimetals. The band crossings in these gapless topological systems occur at distinct points [Fig.~\ref{fig:nodal}(a)] \cite{armitage18rmp,bradlyn16sc,isobe16prb,tang17prl}. At each nodal point, the low-energy theory manifests the chiral fermions with multifold degeneracy [Fig.~\ref{fig:band}(a)]. A minimal model exhibits a rotationally symmetric spin-$s$ Hamiltonian with an integer or half-integer spin $s=1/2,1,3/2,\dots$,
\beeq
\label{eq:cms}
\mca H_{\txt{CMS},\mbf k}=v\mbf k\cdot\mbf S.
\eneq
Here $v$ is the effective velocity, $\mbf k=k\mbf{\hat k}$ with radial magnitude $k$ and direction $\mbf{\hat k}$, and $\mbf S=(S^x,S^y,S^z)$ is the spin-$s$ representation. There are $2s+1$ bands in this model, carrying the dispersion energies $\ve^{sn}_{\mbf k}=vkn$ with $n=-s,-s+1,\dots,s$. In accordance with the Berry flux $\mbf B^{sn}_{\mbf k}=-(n/k^2)\mbf{\hat k}$, the nodal point at $\mbf k=\mbf0$ carries a quantized Berry monopole $q^{sn}=-n$. This monopole charge corresponds to the Chern number $C^{sn}=(1/2\pi)\oint d\mbf S_{\mbf k}\cdot\mbf B^{sn}_{\mbf k}=2q^{sn}=-2n$ under an enclosing-surface integration, which is the topological invariant. The rotationally symmetric structure around the monopole indicates the duality to a Haldane sphere \cite{haldane83prl,jainbook,hsiao20prb}. Analogous to the quantum Hall effect on the Haldane sphere, the ``dual Landau level quantization'' occurs in the trace of quantum metric herein \cite{lin21prb}. An alternative form of the Hamiltonian $\mca H_{\txt{CMS},\mbf k}=vk\mbf{\hat k}\cdot\mbf S$ indicates that the wave functions are independent of $k$. Due to the absence of radial contribution, the trace of quantum metric $\Tr g^{sn}_{\txt{CMS},\mbf k}=\epvl{|\bsb\L^{sn}|^2/k^2}$ measures the dual rotational energy of the dynamical angular momentum $\bsb\L^{sn}=\mbf r^{sn}\times\mbf k$. Note that the angular momentum under rotation symmetry takes the form $\mbf L^{sn}=\bsb\L^{sn}+q^{sn}\mbf{\hat k}$. With $|\bsb\L^{sn}|^2=|\mbf L^{sn}|^2-(q^{sn})^2$, we obtain the quantized trace of quantum metric
\beeq
\Tr g^{sn}_{\txt{CMS},\mbf k}=\fr{1}{k^2}[s(s+1)-(q^{sn})^2],
\eneq
labeled by the angular momentum $s$ and the monopole charge $q^{sn}$. This exemplifies the quantized band geometry from the dual Haldane sphere in the $\mbb Z$ nodal-point semimetals.

The duality can also be adopted to the $\mbb Z$ nodal-surface (NS) semimetals \cite{turker18prb}, which accommodate two-band crossings on closed surfaces [Fig.~\ref{fig:nodal}(b)]. The $\mbb Z$ nodal surfaces may be realized, for example, from a pair of degenerate chiral multifold semimetallic points $\mbf P_\pm$. When the degeneracy is broken by an energy splitting $\D\ve_{\mbf0}=\ve^+_{\mbf0}-\ve^-_{\mbf0}>0$ [Fig.~\ref{fig:band}(b)], the bands from different nodal points cross at spherical nodal surfaces. As inflated nodal points \cite{agterberg17prl}, these nodal surfaces inherit the $\mbb Z$ topological structures of the original nodal points. Consider the low-energy theory of a nodal surface at the energy $\ve_\txt{NS}\in(\ve_{\mbf0}^-,\ve_{\mbf0}^+)$, which is formed by the $n_+$-th and $n_-$-th original bands. For the lower band of the nodal surface, the band eigenstate corresponds to the $n_\mp$-th original band inside(outside) the nodal surface. This implies the occurrence of different Berry fluxes $\mbf B^\txt{in/out}_{\mbf k}=\mbf B^{sn_\mp}_{\mbf k}$ and according Chern numbers $C^\txt{in/out}=C^{sn_\mp}$ in the two regions. A topological invariant is defined by the change of Chern number across the nodal surface $\D C=C^\txt{out}-C^\txt{in}$ \cite{turker18prb}. The quantum metric can also be inferred directly, with the individual components derived for the spin-$1/2$ case in Ref.~\onlinecite{salerno20prr}. Here we present an interpretation based on the duality. The low-energy theory is dual to a free-particle model with a ``uniformly charged magnetic spherical shell'' and a magnetic monopole at the origin. While the monopole charge is $q=C^\txt{in}/2$, the spherical shell carries the total charge $Q=\D C/2$ corresponding to the topological invariant. The Berry charges induce the dual Landau level quantization in the trace of quantum metric
\beeq
\Tr g^\txt{in/out}_{\txt{NS},\mbf k}=\Tr g^{sn_\mp}_{\txt{CMS},\mbf k}.
\eneq
Such a quantization holds generically for the bands near all nodal surfaces. Notably, the band geometry exhibit different structures on different sides of the nodal surfaces. This feature is attributed to the Berry charges at the nodal surfaces and is generic in the $\mbb Z$ nodal-surface semimetals.

\subsection{Berry vortices: Quantization of dual axial rotational energy}

We next turn to the analysis of gapless topological systems with Berry vortices. Before diving into the 3D systems, we discuss how the dual energy quantization occurs at the 2D Dirac points (2DDP). This example will serve as an important hint to the 3D nodal-loop semimetals.

A minimal model at the 2D Dirac points takes a rotationally symmetric form \cite{castroneto09rmp}
\beeq
\mca H_{\txt{2DDP},\mbf k}^l=v(k_+^{2l}\s^-+k_-^{2l}\s^+),
\eneq
with $k_\pm=k_x\pm ik_y$, $l=1/2,1,3/2,\dots$, and Pauli matrices $\s^\pm=(\s^x\pm i\s^y)/2$. This system contains two bands with dispersion energies $\ve^\pm_{\mbf k}=\pm vk^{2l}$, which become degenerate at the Dirac point $\mbf k=\mbf0$. The Berry connection forms a vortex structure around the Dirac point, thereby driving the $\pm2\pi l$ phase windings in the band eigenstates. Accordingly, the Berry flux experiences a singularity at the Dirac point and vanishes everywhere else. The trace of quantum metric is determined by this Berry vortex structure. According to the Hamiltonian $\mca H_{\txt{2DDP},\mbf k}^l=vk^{2l}(\hat k_+^{2l}\s^-+\hat k_-^{2l}\s^+)$, the wave functions are $k$-independent. Without the radial contribution, the trace of quantum metric measures the dual axial rotational energy $\Tr g^{ln}_{\txt{2DDP},\mbf k}=\epvl{(\L^{ln}_z)^2/k^2}$. Here $\L^{ln}_z=L^{ln}_z$ serves as the angular momentum due to the absence of Berry monopole. The magnitude $|L^{ln}_z|=l$ corresponds to the $\pm2\pi l$ phase winding around the Dirac point. We arrive at the dual energy quantization in the trace of quantum metric
\beeq
\Tr g_{\txt{2DDP},\mbf k}^{ln}=\fr{l^2}{k^2},
\eneq
consistent with a direct calculation for $l=1/2$ in graphene \cite{liang17prb}. Our duality-based analysis demonstrates a natural interpretation to the band geometry from the Berry vortices.

The result at the 2D Dirac points can be generalized to the 3D nodal-loop semimetals, where the band crossings take place along closed loops [Fig.~\ref{fig:nodal}(c)] \cite{carter12prb,weng15prx,fang15prb,liang16prb,zhao17prl,bzdusek17prb,ahn18prl}. Consider a minimal two-band model of the nodal-loop semimetals \cite{huh16prb,nandkishore16prb}
\beeq
\mca H_{\txt{NL},\mbf k}(\D)=\fr{k_\perp^2-\D}{2m}\s^x+v_zk_z\s^y,
\eneq
where $k_\perp=(k_x^2+k_y^2)^{1/2}$ and $\D>0$. This system shows a nodal loop in the $k_x$-$k_y$ plane at charge neutrality [Fig.~\ref{fig:band}(c)], which is defined by the major radius $k_R=\sqrt\D$. Assume an isotropic dispersion around the nodal loop $v_\perp=k_R/m=v_z=v$. A linearization at low energy leads to the $l=1/2$ case of the general minimal model \cite{yu19prb,wang20prb}
\beeq
\mca H_{\txt{NL},\mbf k}^l=v(k_{r,+}^{2l}\s^-+k_{r,-}^{2l}\s^+),
\eneq
where $k_{r,\pm}=(k_\perp-k_R)\pm ik_z$ are defined around the nodal loop. This model obeys two rotation symmetries, which are about the tangential axis and the center line $k_\perp=0$ of the nodal loop, respectively. The two bands exhibit the dispersion energies $\ve^\pm_{\mbf k}=\pm vk_r^{2l}$, where $k_r=[(k_\perp-k_R)^2+k_z^2]^{1/2}$ is the minor radius from the nodal loop. Importantly, the model serves as a ``stacking'' of the 2D Dirac-point systems along the nodal loop. This identifies the nodal loop as a Berry vortex line, which is a generalization from the Berry vortex at the 2D Dirac point. The $\pm2\pi l$ phase winding now occurs around the tangential axis of the nodal loop. Accordingly, the Berry flux experiences a singularity along the nodal loop and vanishes everywhere else. The Berry vortex line induces the dual rotational energy of the axial angular momentum $|L^{ln}_\phi|=l$, where $\bsb{\hat\phi}$ is the tangent unit vector along the nodal loop. The trace of quantum metric thus exhibits the dual energy quantization
\beeq
\Tr g_{\txt{NL},\mbf k}^{ln}=\fr{l^2}{k_r^2},
\eneq
similar to the result at the 2D Dirac points. Note that the nodal loop in $\mca H_{\txt{NL},\mbf k}(\D)$ shrinks to a point and vanishes when $\D$ decreases across zero. Such a self annihilation can occur under the combined parity and time-reversal $PT$ symmetry, which imposes the reality condition on the systems. Since the nodal loop can be annihilated by itself, it is $\mbb Z_2$-trivial under $PT$ symmetry in the topological classification \cite{fang15prb}.

There also exist nodal loops which are $\mbb Z_2$-nontrivial under $PT$ symmetry \cite{fang15prb,zhao17prl,bzdusek17prb,ahn18prl}. Consider a minimal model of the $\mbb Z_2$-nontrivial nodal loops \cite{fang15prb}
\beeq
\mca H_{\mbb Z_2\txt{NL},\mbf k}=vk_x\s^x+vk_y\tau^y\s^y+vk_z\s^z+m\tau^x\s^x,
\eneq
where the Pauli matrices $\s^a$ and $\tau^a$ are defined. This model accommodates four bands with the dispersion energies $\ve^{\pm\pm}_{\mbf k}=\pm v[(k_\perp\pm k_R)^2+k_z^2]^{1/2}$ for $k_R=|m|/v$. At finite $|m|>0$, the two middle bands form a nodal loop at charge neutrality with the major radius $k_R$ [Fig.~\ref{fig:band}(d)]. These two bands obey the same rotation symmetries as the model of $\mbb Z_2$-trivial nodal loops. Similar to the $\mbb Z_2$-trivial nodal loop with $l=1/2$, the nodal loop herein serves as a Berry vortex line for a $\pm\pi$ phase winding. An important difference is the interplay with additional nodal lines $k_\perp=0$ at finite energy \cite{ahn18prl}. For the upper (lower) band, an additional nodal line occurs at the band crossing with the highest (lowest) band. This additional nodal line penetrates the center of the nodal loop, leading to a topologically nontrivial ``link'' [Fig.~\ref{fig:nodal}(d)]. The link with the additional nodal line defines a $\mbb Z_2$ monopole at the nodal loop. This monopole drives the nodal loop $\mbb Z_2$-nontrivial, which shrinks to a point (when $m=0$) at most under changing $m$. The $\mbb Z_2$-nontrivial nodal loops can only be annihilated through the pair annihilation. Such an annihilation may occur, for example, between two nodal loops which are linked to the same nodal line.

The link of the $\mbb Z_2$ monopole is encoded in the band geometry. We calculate the quantum metric for the two middle bands in the minimal model. The individual components of quantum metric are complicated, with the special cases derived at $k_z=0$ in Ref.~\onlinecite{salerno20prr}. Nevertheless, the trace of quantum metric is amiable according to the duality. It can be shown that the wave functions are independent of the minor and major radial components $k_r$ and $k_R$. This indicates the absence of radial contributions in the trace of quantum metric, leaving only the angular contributions. We further uncover that each wave function is a tensor product of two $\mbb Z_2$-trivial wave functions, which are related to the nodal loop and the additional nodal line, respectively. A direct calculation obtains the according dual axial rotational energy in the trace of quantum metric
\beeq
\Tr g_{\mbb Z_2\txt{NL},\mbf k}=\fr{1}{4k_r^2}+\fr{1}{4k_\perp^2}.
\eneq
The first term is the dual axial rotational energy from the nodal loop, which is identical to the $\mbb Z_2$-trivial case with $l=1/2$. Meanwhile, the second term captures the dual rotational energy of the axial angular momentum $|L^n_z|=1/2$ about the additional nodal line. This additional term clearly indicates the link of the $\mbb Z_2$ monopole, which is absent for the $\mbb Z_2$-trivial nodal loop. We thus arrive at a clear interpretation of the nontrivial band geometry in the $\mbb Z_2$ nodal-loop semimetals.

\section{Practical manifestations}

We identified various types of dual energy quantization in the band geometry. An important question is whether these results can be observed in the practical systems. A direct manifestation of the band geometry occurs in the spread of Wannier functions \cite{strinati78prb,marzari97prb,matsuura10prb,marzari12rmp,lee14prb,lin21prb}. For the Wannier functions of a band, the spread functional $\O=\epvl{|\gradk|^2}-|\epvl{\gradk}|^2$ exhibits the gauge-invariant lower bound $\O_\txt{GI}=\mca V_0\Sigma$ with $\Sigma=\intv{k}\Tr g_{\mbf k}$. Here $\mca V_0$ is the unit-cell volume and $\intv{k}=\int d^dk/(2\pi)^d$ with $d=2,3$ is the integral over the Brillouin zone.  As the integrated trace of quantum metric, the gauge-invariant lower bound originates solely from the band geometry. Since the trace of quantum metric is positive-semidefinite, a further estimation $\Sigma_\txt{SR}=\int_{\txt{SR},\mbf k}\Tr g_{\mbf k}\leq\Sigma$ can be obtained by the focus on a specific subregion (SR). For the gapless topological systems, the trace of quantum metric reaches the maxima at the band crossings with Berry defects. The subregions near the band crossings thus support good estimations to the lower bound.

For the 3D chiral multifold semimetals, the estimated lower bound is quantized \cite{lin21prb}
\beeq
\Sigma_\txt{CMF}^{sn}=\fr{\mca V_0\L_k}{4\pi^2}2[s(s+1)-(q^{sn})^2].
\eneq
Here $\L_k$ is an ultraviolet (UV) radial cutoff in the focused subregion. The quantized result also applies to the other gapless topological systems with Berry charges, such as the $\mbb Z$ nodal-surface semimetals. Significantly different features are manifest from the Berry vortices. For the 2D Dirac points, the estimated lower bound is logarithmically divergent \cite{thonhauser06prb,ma10prb}
\beeq
\Sigma_\txt{2DDP}^{ln}=\fr{\mca V_0}{2\pi}l^2\ln k|_0^{\L_k}.
\eneq
Such a logarithmic divergence is naturally generalized to the $\mbb Z_2$-trivial nodal loops
\beeq
\Sigma_\txt{NL}^{ln}=\fr{\mca V_0k_R}{2\pi}l^2\ln k_r|_0^{\L_{k_r}}.
\eneq
The $\mbb Z_2$-nontrivial nodal loops exhibit additional logarithmic divergences
\beeq
\Sigma_{\mbb Z_2\txt{NL}}-\Sigma_\txt{NL}^{(1/2)(\pm1/2)}\sim\ln k_\perp|_{\L_{k_R}^\txt{min}}^{\L_{k_R}^\txt{max}}
\eneq
from the additional nodal lines. This distinguishes the estimated lower bound of the $\mbb Z_2$-nontrivial nodal loop from the $\mbb Z_2$-trivial one. Due to the logarithmic divergences, the band geometry may have strong manifestations at the band crossings with Berry vortices. Note that the full sets of bands are still Wannierizable with finite spreads \cite{brouder07prl}. The seeming contradiction can be resolved by choosing the ``smooth gauges'' for the full sets of bands \cite{soluyanov12prb}. Under the transformation to a smooth gauge, the singular Bloch functions in the band basis are transformed into regular functions in the whole Brillouin zone. The finite spreads can then be yielded from these regular Bloch functions.

In the practical systems, the rotation symmetries may be broken under lower crystalline symmetries or away from the band crossings. Under the breakdown of rotation symmetries, the simple quantization rules may be broken. Nevertheless, the results in our analysis may remain close to the quantized values until the band crossing structures completely change. Consider the integrated trace of quantum metric at a fixed radius $k$ in a 3D $\mbb Z$ nodal point or 2D Dirac point. When a perturbation breaks the rotation symmetry, the quantized Berry charge or Berry vortex may still pull the nonquantized result close to the original quantization. A stronger symmetry breaking may push the system across a topological phase transition, where the Berry charge or Berry vortex changes. The integrated trace of quantum metric may become divergent across the topological phase transition and fall into the vicinity of another quantized value \cite{thonhauser06prb,ma10prb}. Similar effects can also occur at the 3D nodal loops and nodal surfaces.

The gauge-invariant lower bound can be probed directly with a periodic drive \cite{ozawa18prb}. Under a linear shake along a direction $\mbf{\hat a}$, the integrated quantum metric $\intv{k}g_{aa\mbf k}$ of the occupied bands corresponds to the excitation rate. The integrated trace of quantum metric is then obtained from the measurements along all directions, which determines the gauge-invariant lower bound. Such a measurement was performed experimentally in an ultracold atomic Floquet Chern insulator \cite{asteria19np}. Most of the current measurements apply to the integrated trace of quantum metric of all occupied bands. The probe with momentum and band resolutions will advance the understanding of band geometry in the gapless topological systems \cite{ozawa18prb,aidelsburger15np,chen20ax}.

\section{Discussion}

We showed that the position-momentum duality offers a transparent interpretation of the band geometry at the topological band crossings. The trace of quantum metric provides a unified framework through this duality, which is achieved by encoding the dual rotational energy about the band crossings with nontrivial Berry defects. This scenario applies to the Berry charges at the 3D $\mbb Z$ nodal points and nodal surfaces, as well as the Berry vortices at the 2D Dirac points, 3D $\mbb Z_2$-trivial and nontrivial nodal loops. Our duality-based analysis offers a feasible route toward further understanding of nontrivial band geometry. The investigations in the other gapless topological systems, such as the gapless topological superconductors \cite{sigrist91rmp,nandkishore16prb,agterberg17prl,brydon18prb,venderbos18prx,link20prl} and spin liquids \cite{savary16rpp}, degenerate bands with non-Abelian Berry connections \cite{ma10prb}, or higher-dimensional systems with tensor monopoles \cite{palumbo18prl,chen20ax}, may serve as interesting topics for future work.

\begin{acknowledgments}
The authors thank Nathan Goldman, Ching Hua Lee, and especially Rahul Nandkishore for fruitful discussions and feedback on the manuscript. YPL was sponsored by the Army Research Office under Grant No. W911NF-17-1-0482. The views and conclusions contained in this document are those of the authors and should not be interpreted as representing the official policies, either expressed or implied, of the Army Research Office or the U.S. Government. The U.S. Government is authorized to reproduce and distribute reprints for Government purposes notwithstanding any copyright notation herein. YPL also acknowledges Ying-Jer Kao for the hospitality at National Taiwan University during the pandemic.
\end{acknowledgments}



\bibliography{Reference}

\end{document}